# ELECTROMECHANICAL STUDY OF A RING-BRUSH SLIDING CONTACT

E. Chevallier[a,*], T. Garcia[b], and S. Ait Mohamed[c]

*eddy.chevallier@univ-reims.fr

[a] Institut de Thermique, Mécanique, Matériaux, (ITheMM – UR URCA 7548), Université de Reims Champagne-Ardenne, Campus du Moulin de la Housse, 51687 Reims, France

[b] Laboratoire « Physique des Systèmes Complexes » (PSC – UR UPJV 4663), Université de Picardie Jules Verne, 33 rue Saint Leu, 80039 Amiens, France

[c] Laboratoire de Physique de la Matière Condensée (LPMC – UR UPJV 2081), Université de Picardie Jules Verne, 33 rue Saint Leu, 80039 Amiens, France

**ABSTRACT**

We report a study about the electrical response from a sliding contact made of a silver-graphite brush and a brass ring. This study focuses specifically on the voltage variations due to the mechanical interactions across the contact according to the rotational speed. This study is part of the research and the development about the monitoring of dynamical interfaces.

## 1. INTRODUCTION

In some devices such as a rotary current collector the transfer of power currents or any signals is carried out through a sliding contact, often established between a conductive brush, or a metallic wire, and a metallic ring. However, signals sent through such a sliding contact always emerge with a noisy component, leading to the jamming of signals. Such noise comes from the electromechanical interactions (coupling) between contacting asperities at the interface [1, 2].

The electromechanical characterization of electrical contacts and its associated difficulties are not that clear [3-7]. Despite this, the use of voltage noise across the contact as a mean to probe the surface state has been successful for over a decade [8-17]. As instance, a complete study of the long-term evolution of the main electrical properties of a gold-coated ring-wire sliding contact was earlier reported [8]. This study based on measurable relevant quantities such as the voltage noise revealed the ageing of the contact. Subsequently, a numerical study was made on the electrical response from a mechanical model describing the contact disturbances versus sliding speed [9]. This leads to the improvement of a monitoring method in real time and very sensitive of contacts quality.

The present work focusses on the description of the contact disturbances between a silver-graphite brush and a brass ring versus the rotational speed. The ring-brush contact is particularly complex: the pressure exerted on the brush has to be optimal [18, 20]. It must be strong enough to ensure continuous contact on the ring under all working conditions, and so to avoid arcing problems [21, 22], and must be weak enough to not increase friction (heat) and wear. The optimum brush pressure is got from the combination of both electrical and mechanical considerations: the suitable pressure is thus a compromise between the wear processes of each of these physical phenomena (Fig.1) - which are also related to the contact heating [23-27]. Silver-graphite brushes have a higher conductivity than copper-graphite ones and form a special low resistance film due to the conductivity of silver oxide. Moreover, silver grade brushes are also known to be able to transfer low voltage current signals without degradation.

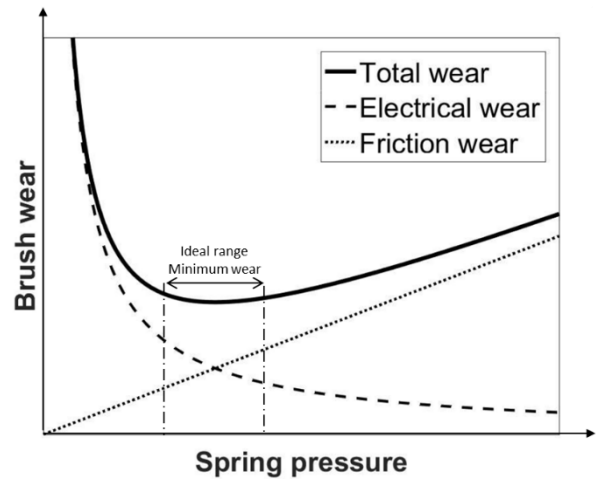

Fig. 1. Wear evolution as function of the contact pressure of the brush.

Ring-brush contact has another particularity: the heterogeneity of contacting materials leads to the polarization of the system. At the cathode the oxidation is favored and the electric field leads metallic ions into the third body [28]. At the anode the graphite deposit is favored and the voltage fluctuations are very weak or null [29-32]. Also at the cathode, the contact resistance is higher [4]. Moreover, the polarization has influence on the friction wear: since the anode favors the deposit of graphite, hence the wear track is lubricated by the brush [28] leading to a correlation between electrical and tribological properties [33].

Therefore, our study follows two steps: (i) get the pertinent observables through measurements versus rotational speed and (ii) model the electromechanical coupling to emphasize the interpretation and the meaning of experimental data. The model, based on a phenomenological description of the mechanical interactions, combines the tribological aspects of the sliding contact to the electrical ones and lies on two main assumptions: (i) the contact spots are mechanically and electrically independent and (ii) the contact number and its fluctuation are respectively piloted by the contact load $F$, i.e. the contact pressure, which tends to maintain the surfaces in contact, and the rotational speed $\omega$.

Note that, despite the previous identification of additional contributions to the electrical contact noise as chemical inhomogeneity [8], heating, or any tribo-films, the present work is only focus on the exploration of the mechanical influences.

## 2. EXPERIMENTAL SETUP AND RESULTS

2.1 Materials

The experimental setup, shown on figure 2, is composed of a spindle of five rotating brass rings. Each ring has a diameter of about 6 cm, and the spindle have a misalignment of less than 1 mm. On each ring slides dryly a pair of brushes, with a mass of 1.5 g each, that are pressed on the ring with a force of about 1 N by a spring with a stiffness of about 200 N/m. Due to the geometry of the contacting elements and the effect of the rotation, the contacting zone is smaller than those given initially by the brush – about 1/3 of the total area – and is estimated to about 8 mm² (Fig.3a). Figure 3b gives an image of the contact surface after lapping which will be representative of its state during measurements.

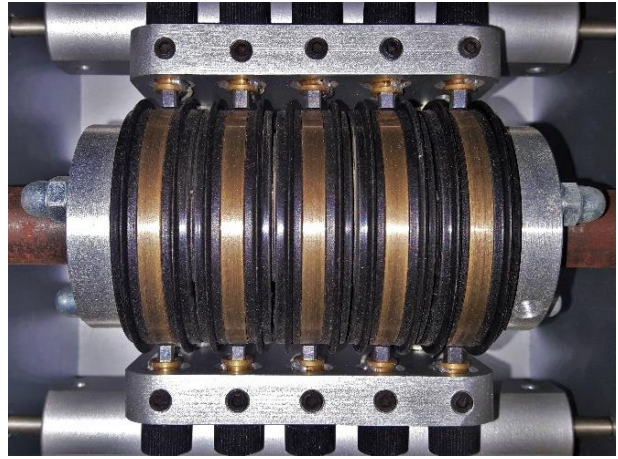

Fig. 2. Experimental bench.

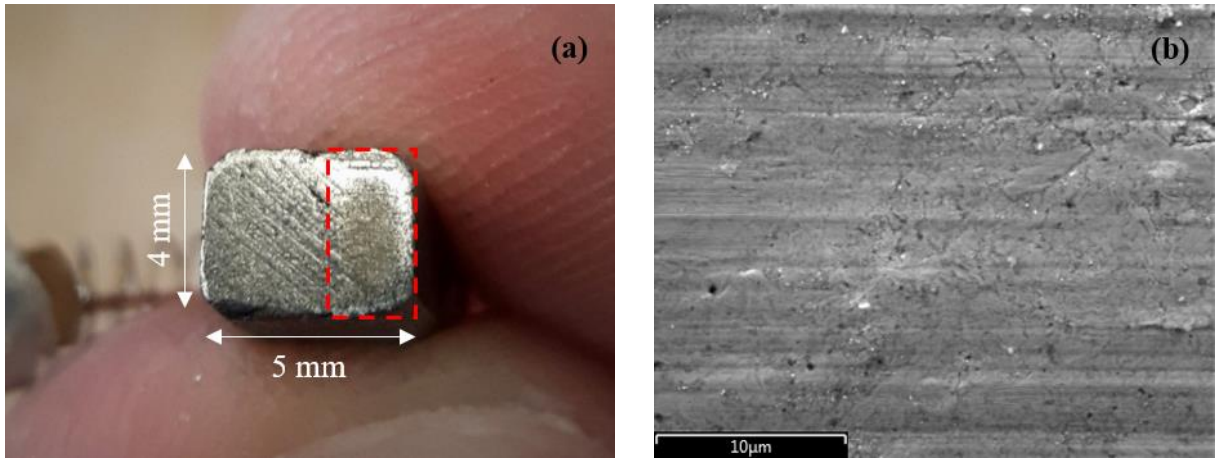

Fig. 3. (a) Contact zone of the brush (dotted square). (b) SEM micrograph of the contact zone, free of arcing marks.

The chemical composition of the contact area was characterized using a scanning electronic microscope - Quanta 200FEG, with a microanalyzer X INCA Oxford type SDD 80 mm². Such composition was acquired at high vacuum and 10 KeV, and is given in Table 1. Images of the chemical distribution of each detected element are shown is figure 4. The main chemical components of the brush were silver (73 %w), copper (14 %w) and carbon (7 %w).

| Element | % | $\sigma_\%$ |
|---|---|---|
| Ag | 73.24 | 0.18 |
| Cu | 14.20 | 0.11 |
| C | 6.73 | 0.07 |
| O | 4.55 | 0.08 |

Table 1. Chemical composition of brushes.

2.2 Protocol

In each pair of brushes flowed an imposed DC current, $I$, equal to 200 mA, and the voltage across them, $V$, was recorded as a function of time using a Keithley SourceMeter 2611A using the 4 points method at a sampling of 10 points/s in a window of 100s (1000 points). The recording was made for different rotational speeds, $\omega_r$, on each of the five rings. The value of the probing current (200 mA) was chosen according a primary analysis that showed that the contact resistance was independent of the applied current when this one is between 200 mA and 500 mA.

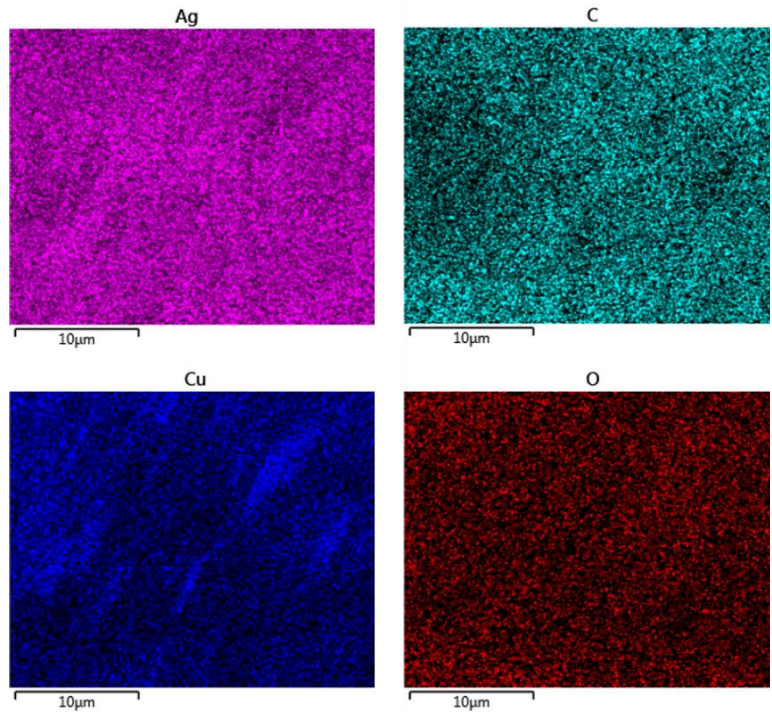

Fig. 4. Chemical distribution of the contact zone.

2.3 Experimental results

Figure 5 shows the electrical equivalent of the system and highlights what is accessed through the measurement: $Z_w$ is the contribution of wires, $Z_r$ the contribution of the ring bulk, and both $Z_c$ and $Z'_c$ are the contribution of interface phenomena.

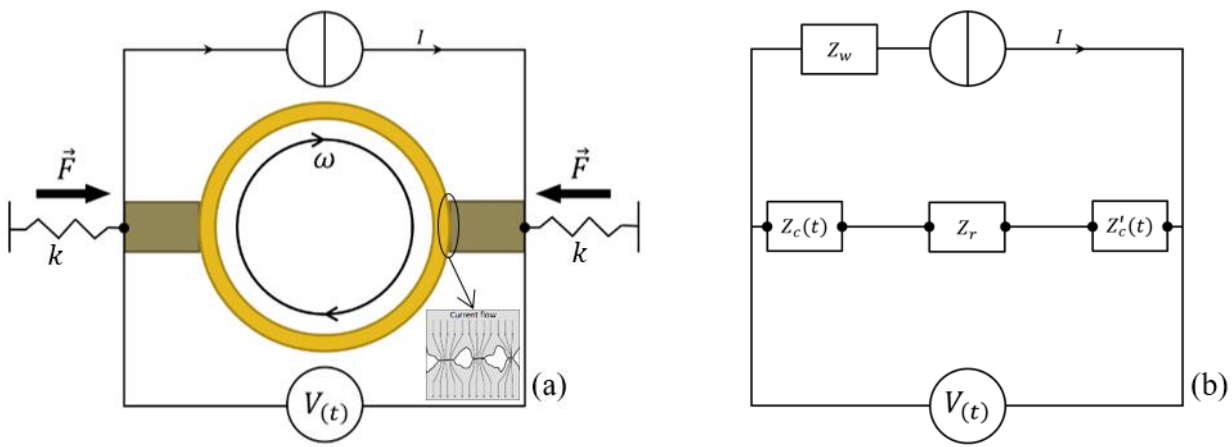

Fig. 5. (a) Scheme of the experimental setup, and (b) its electrical equivalent.

Figure 6 shows a typical measurement (a) and its associated Fourier transform (b). Figure 7 shows the evolution of the average value, $A_0$, of a measurement (null frequency component) versus the rotation speed. Data are fitted by two complementary approaches: the first one treats separately the two behaviors (fit1 and fit2 – linear laws) and the second treats globally the evolution (fit3 – power law). Figure 8 shows the evolution of the main frequency, $A_1$, and its third first harmonics, $A_2$, $A_3$, and $A_4$. Figure 9 shows the evolution of the average power spectral density (PSD) of the DC component of the spectrum.

Note that the mains values and their respective error bars are calculated from a statistical operation over the set of the five rings.

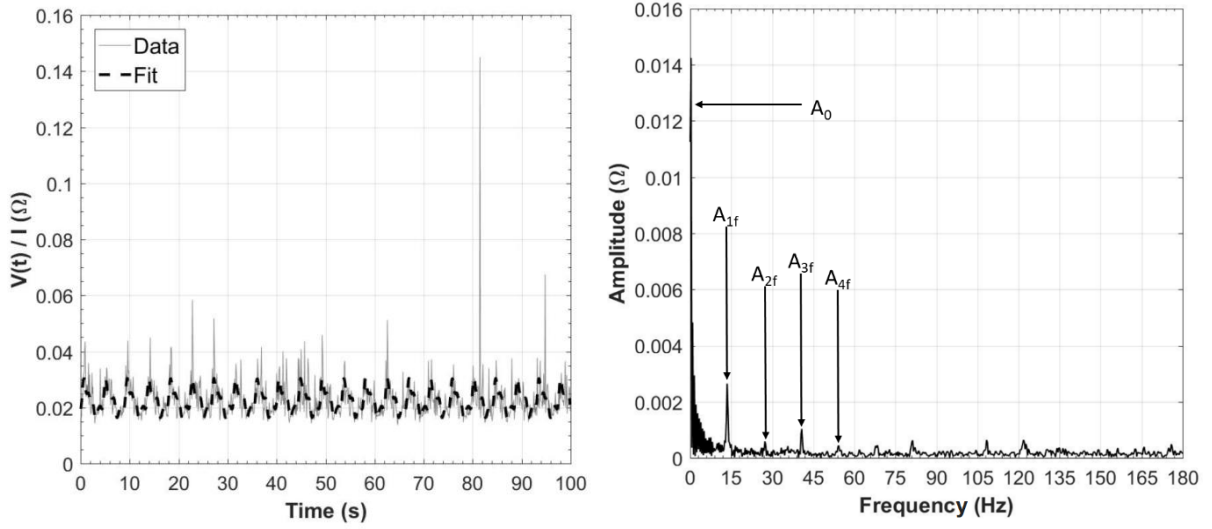

Fig. 6. (a) Ratio $V(t)/I$ for $I$ = 200 mA and $\omega$ = 13.69 rpm, fitted using the "cftool" toolbox from MatLab R2018b, and (b) its Fourier transform.

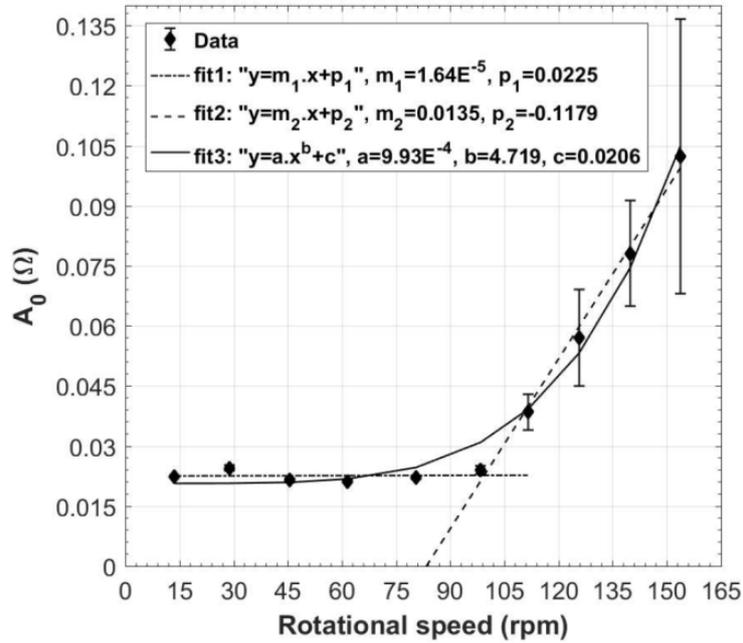

Fig. 7. Evolution of the null frequency component, $A_0$, versus the rotational speed.

## 3. ELECTROMECHANICAL MODELING

To model the behavior of the ring-brush contact we need to distinguish two complementary mechanical contributions: the macroscopic ones, i.e. the brush displacement along its slot – normal to the surface, and the microscopic ones, i.e. the contact noise (asperities interactions) due to the relative displacement – tangent to the surface.

3.1 Macroscopic scale

As shown in figure 5a, each brush is flattened on the ring with a force $\vec{F}$ from a spring of stiffness $k$. Due to the ring waviness and/or misalignment, brushes are displaced periodically around a mean position during the rotation: the displacement leads to a compression or a dilation of their respective spring and therefore leads to module the force applied on their associated contact.

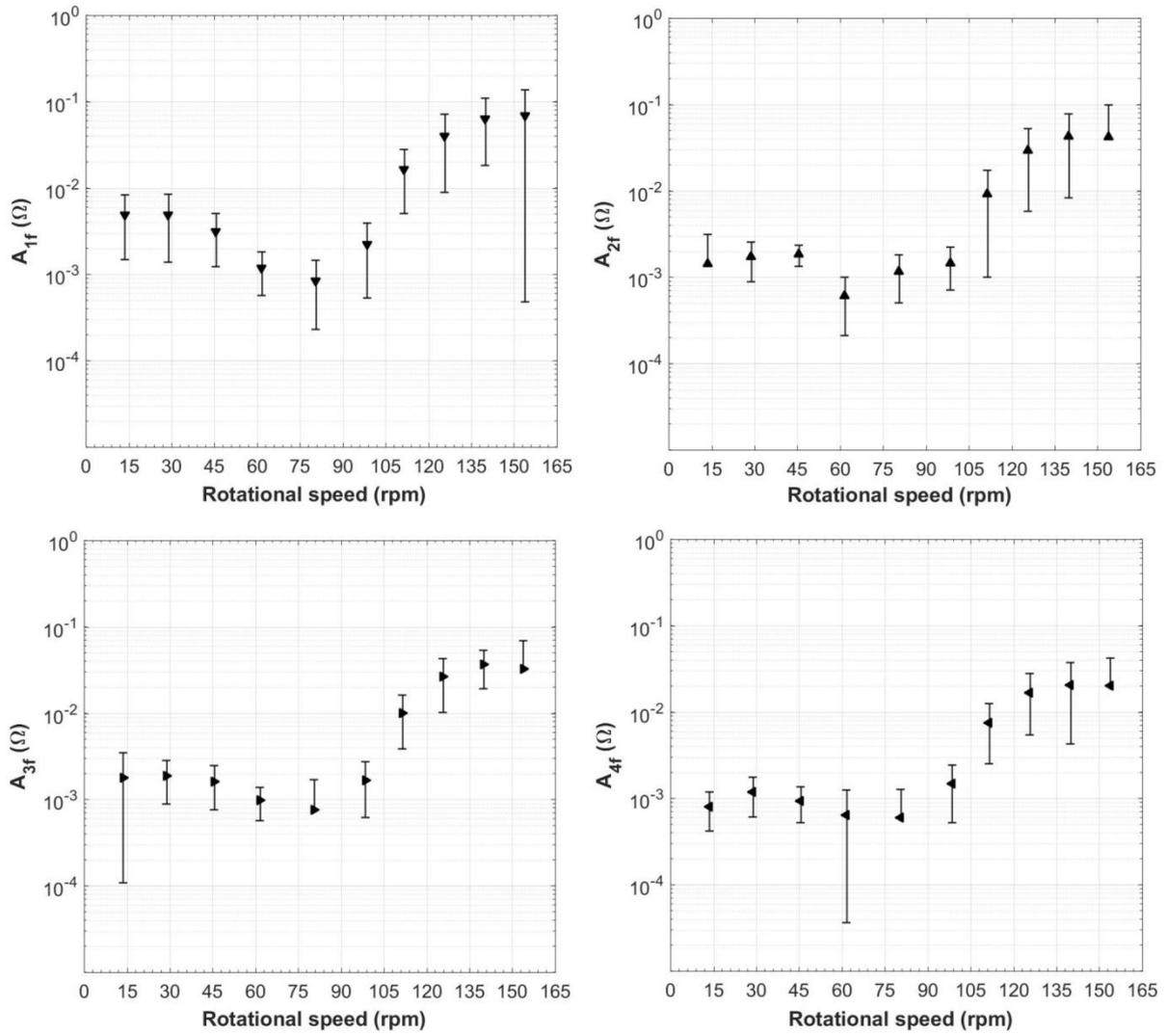

Fig. 8. Evolution of the four first harmonics versus rotational speed.

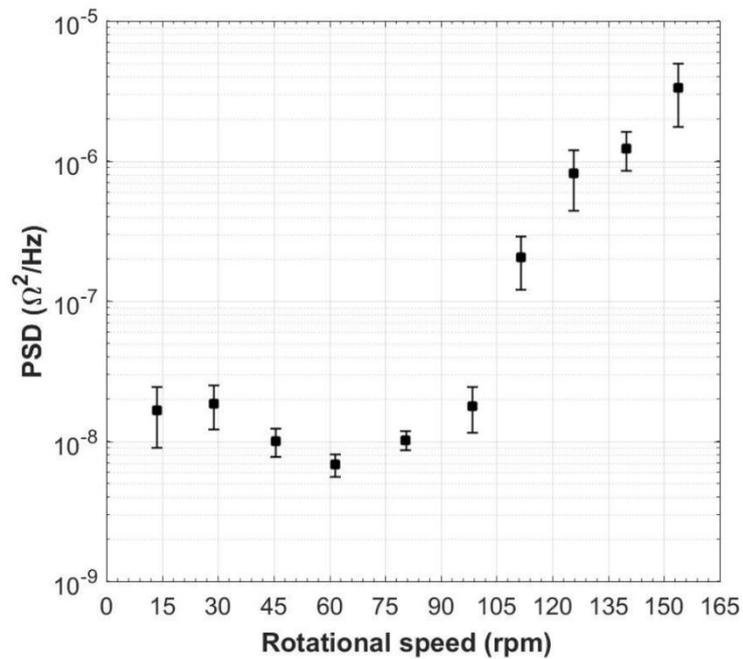

Fig. 9. Evolution of the average DC component of the power spectral density (PSD), versus the rotational speed.

Restricting to a 1D problem by assuming that the motion of the brush holds within the applied force axis, we can advantageously model the displacement as the space fluctuations of the ring radius $\delta_R$, which is a periodic function of the angle $\theta$ along the ring: $\delta_R = f(\theta)$. This displacement is accompanied by its own acceleration, $\ddot{\delta}_R$, that appears in the system as an inertial force, $\vec{F}_i$, related to the rotational speed, $\omega_r = d\theta/dt$, and related to the surface state. Indeed, since $\delta_R = f(\theta(t))$, it yields:

$$\ddot{\delta}_R = \frac{d^2 f}{dt^2} = \frac{d^2 f}{d\theta^2} \cdot \omega_r^2 + \frac{df}{d\theta} \cdot \frac{d\omega_r}{dt} \qquad [1]$$

where $d^2 f/d\theta^2$ is the term directly related to the surface curvature of the ring (topography).

During the rotation, the brush and the ring remain in contact. Thus, making the balance of the forces applied on the system, neglecting the friction of the brush in its slot, it yields:

$$\vec{F} + \vec{F}_i + \vec{R}_n = \vec{0} \qquad [2]$$

which, once detailed, gives:

$$-k \cdot \delta_R - F_0 - m \cdot \ddot{\delta}_R + R_n = 0 \qquad [3]$$

and finally, for a constant rotational speed, we have:

$$R_n = k \cdot f(\theta(t)) + m \cdot \frac{d^2 f(\theta(t))}{d\theta^2} \cdot \omega_r^2 + F_0 \qquad [4]$$

where $m$ is the mass of the brush and $F_0$ the constant force around which the modulation is made: $\langle F \rangle_\theta = F_0$.

Eq. [4] highlights through the expression of the normal reaction, $R_n$, how the contact undergoes the modulation of the applied force. For low rotational speeds, the contact undergoes a load directly proportional to the force applied by the spring. For higher rotational speeds, according to the sign of $d^2 \delta_R/d\theta^2$, the contact undergoes a load that can significantly increase, decrease or even be null. A null, or negative, normal reaction means that the contact between the brush and the ring is open – and leads consequently in this context to an interruption of the current flow. Eq. [4] therefore indicates the condition of stability of the ring-brush contact.

3.2 Microscopic scale

The contact between the ring and a brush is a multi-contact interface [34]. During sliding, asperities from the brush and the ring create successively temporary junctions through which the current flows. Thus, from an electric point of view, the contact interface can be interpreted as a large circuit of parallel impedances, where each one has a switch, $S(t)$, that open ($S = 0$) and close ($S = 1$) by following the influence of the surface state (roughness), the applied force, and of the sliding speed (Fig. 10). The mathematical expression describing the switch function for the $i^{th}$ contact in a given interface is presented by Eq. [5], where $H$ is the function of Heaviside, and $t_j$ the instant when the shock happen.

$$S_i(t) = \sum_j H(t - t_j) - H(t - t_j - t_c) \qquad [5]$$

The next shocking instant is defined by: $t_{j+1} = t_j + t_c + t_{\bar{c}}$, where $t_c$ and $t_{\bar{c}}$, are respectively the contact duration and the no-contact duration, which are distributed each around a mean value, fixed by mechanical parameters, that remains to be determined.

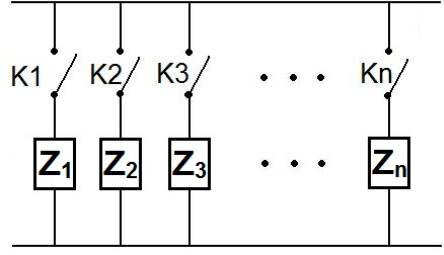

Fig. 10. Scheme of the multi-contact interface from an electrical point of view.

Le Bot [2] estimated the colliding rate, $n_s$, for a metal/metal interface based on the work of Rabinowicz [1] founding that is directly proportional to the sliding speed, $v$:

$$n_s = \frac{N_0}{L_c} \cdot v \quad [6]$$

where $N_0$ is the assumed initial contacts number in the interface, and $L_c$ a characteristic length relative to the surface state and which ranges from 1 to 10 µm. The mean contact durations, $t_c$, and the no-contact durations, $t_{\bar{c}}$, can then be estimated as respectively described in Eq. [7] and Eq. [8], where $D$ is the average diameter of a contact spot. These two parameters can thus be estimated according to the distribution of contact diameters ($\bar{D}_\iota$).

$$t_c = 2\bar{D}_\iota/v \quad [7]$$

$$t_{\bar{c}} = (L_c - 2\bar{D}_\iota)/v \quad [8]$$

The diameter of contact spots is a sensitive parameter both related to the surface state and to the applied load: it follows the law of contact defined by Eq. [9] which remains true for ring-brush contact [31, 34], where $0 < \alpha < 1$ and $K$ is a parameter gathering the average mechanical and the geometrical properties of coupling asperities.

$$D = K \cdot F^{\alpha/2} \quad [9]$$

By combining Eq. [7] and Eq. [9] we obtain Eq. [10], showing us the antagonist role between the load and the sliding speed: the contact is better at slow sliding speed and under high load. Moreover, the statistical distribution of durations, $t_c$ and $t_{\bar{c}}$, is related to those of $D$ through the distribution of the mechanical and geometrical parameters gathered in $K'$.

$$t_c = K' \cdot F^{\alpha/2}/v \quad [10]$$

3.3 Electrical considerations

According to the definition of Holm, the contact resistance, $R_H$, for a heterogeneous micro-contact is thus defined by Eq. [11], where $\rho_A$ and $\rho_B$ are the resistivity of each material.

$$R_H = \frac{(\rho_A + \rho_B)}{2K} \cdot F^{-\alpha/2} \quad [11]$$

During sliding, the load applied on the contact will varies according to the modulation of the force described previously at macroscopic scale. Furthermore, the condition of existence of the contact will follow the description at the microscopic scale through the switch function, where durations are directly related to the applied load. Finally, it yields:

$$\frac{1}{R(t)} = \frac{2}{(\rho_A + \rho_B)} \cdot \left(\frac{R_n(t)}{N(t)}\right)^{\alpha/2} \cdot \sum_{i=1}^{N_0} K_i \cdot S_i(t, R_n(t)) \quad [12]$$

**CONCLUSION**

The processing of the measurements leads to identify an operating regime for our system. Beyond 100 rpm the system undergoes disturbances which significantly increase both the average value and the standard deviation of the measured voltage. Processing the power spectral density helps to reinforce this observation. However, we note for the amplitudes of the harmonics, just as for the PSD, that a minimum seems to be reached between 60 and 90 rpm. The speed does indeed have a notable influence on the contact impedance. It remains to consider additional measurements on the spectral properties in order to reveal a bandwidth of the contact (transfer function). This will allow us to know and to understand the limits of use of this kind of system for the transfer of information or control signals.

The simple - but relevant - model established leads to highlight an important parameter at the origin of contact losses during rotation: the geometry of the surface on which the brush slides. The slightest disturbance in the geometry, due to misalignment or local deformation of the ring, can cause contact loses depending on the speed regime. This first model makes possible to estimate the degree of precision required when installing the system or when machining the parts in contact. However, this model needs to be reinforced and completed, particularly at the microscopic scale by considering more the production of the third body (patina) which contributes to the lubrication of the contact and at the same time to the improvement of current flow conditions.

**ACKNOWLEDGMENTS**

Thanks to the electron microscopy platform "Hub de l'énergie" for making his SEM available for SEM and EDX analysis. The authors of this article acknowledge Olivier Bernard from the Mersen Group for discussion and authorization, and for providing the experimental materials.

**REFERENCES**

[1] Rabinowicz, E. (1951), "The Nature of the Static and the Kinetic Coefficients of Friction," Journal of Applied Physics, 22, pp 1373–1379.

[2] Le Bot, A. (2017), "Noise of Sliding Rough Contact," Journal of Physics: Conference Series, 797, pp 012006.

[3] Holm, R. (1967), Electric Contacts: Theory and Applications, 4th ed., Springer, Berlin/Heidelberg.

[4] Féchant, L. (1995), "Le Contact Electrique - L'appareillage de connexion," Hermes Science Publication, Paris, France. ISBN 2-86601-473-1.

[5] Bryant, M. D. (2005), "Tribology Issues in Electrical Contacts," Mechanical Engineering, The University of Texas at Austin, Austin, Texas, 78712-1063.

[6] Schneegans, O., Chrétien, P., Houzé, F. (2011), "Apparatus for measuring the local electrical resistance of a surface." Patent No. PCT/IB2011/051951.

[7] Yastrebov, V. A., Cailletaud, G., Proudhon, H., Mballa, F. S. M., Noël, S., Testé, P., and Houzé, F. (2015), "Three-level Multi-scale Modeling of Electrical Contacts sensitivity study and experimental validation," Proc. 61st IEEE Holm Conference on Electrical Contacts, San Diego CA, USA, October 2015, pp 414-422.

[8] Chevallier, E., Bourny, V., Bouzerar, R., Fortin, J., Durand-Drouhin, O., and Da Ros, V. (2014), "Voltage Noise across a Metal/Metal Sliding Contact as a Probe of the Surface State," Journal of Applied Physics, 115, pp 154903.

[9] Chevallier, E. (2020), "Mechanical Model of the Electrical Response from a Ring–Wire Sliding Contact," Tribology Transactions, Vol. 63, No. 2, pp 215-221.

[10] Machado, C., Baudon, S., Guessasma, M., Bourny, V., Fortin, J., Bouzerar, R., and Maier, P. (2019), "An Original DEM Bearing Model with Electromechanical Coupling," International Journal of Computational Methods, Vol. 16, No. 05, pp 1840006.

[11] Bourbatache, K., Guessasma, M., Bellenger, E., Bourny, V. and Fortin, J. (2013), "DEM ball bearing model and defect diagnosis by electrical measurement," Mechanical Systems and Signal Processing, Vol. 41, No. 1-2, pp 98-112.

[12] Guessasma, M., Bourny, V., Haddad, H., Machado, C., Chevallier, E., Tekaya, A., Leclerc, W., Bouzerar, R., Bourbatache, K., Pélegris, C., Bellenger, E. and Fortin, J. (2018), "Multi-Scale and Multi-Physics Modeling of the Contact Interface Using DEM and Coupled DEM-FEM Approach," Advances in Multi-Physics and Multi-Scale Couplings in Geo-Environmental Mechanics, Elsevier, pp 1-31. ISBN 9781785482786.

[13] Bouzerar, R., Bourny, V., Capitaine, T. and Senlis, J. (2009), "Characterization of the electrodynamic properties of a ceramic-based ball bearing by impedance spectroscopy," Studies in Applied Electromagnetics and Mechanics, Vol. 34, pp 842 - 850.

[14] Jonckheere, B., Bouzerar, R., Bourny, V., Bausseron, T., Foy, N., Durand-Drouhin, O., Le Marrec, F. and Chevallier, E. (2017), "Assessment of the real contact area of a multi-contact interface from electrical measurements," Proc. 23ème Congrès Français de Mécanique (CFM), Lille, France, August 2017.

[15] Bourny, V., Bouzerar, R., Capitaine, T., Fortin, J. and Lorthois, A. (2015), "Surveillance de l'état d'un dispositif de roulement," - *"Monitoring of bearings state"*. Patent No. FR3016041A1.

[16] Bourny, V., Fortin, J., Capitaine, T., Lorthois, A. and Da Ros, V. (2016), "Monitoring the quality of a communication channel supported by a sliding contact." Patent No. US20160142162 (US9917664B2).

[17] Puille, C., Durand-Drouhin, O., Le Marrec, F., Bouzerar, R., Bourny, V., Lejeune, M., Fortin, J., Cantaluppi-Harlée, A., and Andasmas, M. (2018), "Surveillance de l'état d'un dispositif de garniture mécanique," - *"Monitoring of mechanical seals"*. Patent No. FR3065286A1.

[18] Yasar, I., Canakci, A., Arslan, F. (2007), "The effect of brush spring pressure on the wear behaviour of copper–graphite brushes with electrical current," Tribology International, Vol. 40, No. 9, pp 1381-1386.

[19] Koenitzer J. D. (2008), "The Effect of Spring Pressure on Carbon Brush Wear Rate," Helwig Carbon Products, Inc. Milwaukee, Wisconsin, USA.

[20] TR61015 (1990), "Brush-holders for electrical machines - Guide to the measurement of the static thrust applied to brushes," International Electro-technical Commission.

[21] Kubo, S., and Kato, K. (1998), "Effect of arc discharge on wear rate of Cu-impregnated carbon strip in unlubricated sliding against Cu trolley under electric current," Wear, Vol. 216, No. 2, pp 172-178.

[22] Jonckheere, B., Bouzerar, R., Ait Mohamed, S., and Bausseron, T. (2022), "Electrical arc transfer in a multi-contact interface," Sensors and Actuators A: Physical, Vol. 335, pp 113215.

[23] Liu, H.P., Carnes, R.W., Gully, J.H. (1993), "Effect of temperature on wear rate of homopolar pulse consolidated electrical brushes," Wear, Vol. 167, pp 41-47.

[24] Senouci, A., Frene, C., Zaidi, H. (1999), "Wear mechanism in graphite-copper electrical sliding contact," Wear, Vol. 225–229, part 2, pp 949-953.

[25] Dow, T. A., Kannel, J. W. (1982), "Thermomechanical effects in high current density electrical slip rings," Wear, Vol. 79, No. 1, pp 93-105.

[26] Burton, RA. (1985), "Heating and cooling of monolithic current collectors," Proc. current collector conference, Raleigh, NC, USA, pp 299–321.

[27] Zhao, H., Barber, G. C., Liu, J. (2001), "Friction and wear in high speed sliding with and without electrical current," Wear, Vol. 249, pp 409-414.

[28] Godet, M. (1984), "The Third Body Approach: A Mechanical View of Wear," Wear, Vol. 100, pp 437-452.

[29] Shobert, I. E. (1965) "Carbon Brushes: The Physics and Chemistry of Sliding Contacts," Chemical Publishing Company, Inc. New York.

[30] Thomson, J. E. and Turner, J. B. (1962), "The Part Played by a Metal Oxide in Determining the Characteristics of a Graphite Metal Interface," Wear, Vol. 6.

[31] Casstevens, J. M. (1977), "Influence of High Velocities and High Current on the Friction and Wear Behaviour of Copper-Graphite Brushes," Wear, Vol. 48.

[32] Chazalon, P. (2013), "Etudes des variations de la résistance électrique du contact Balai-Bague de l'alternateur", Doctoral Thesis 2013, École Centrale Paris, LISMMA, France.

[33] Mouadji, Y., Bouchoucha, A., Amokrane, B., and Zaidia, H. (2014). "Influence du courant électrique sur le frottement des couples cuivre-graphite, bronze-graphite et graphite-graphite," Proc. 26es Journées Internationales Francophones de Tribologie (JIFT), Mulhouse, France, May 2014. Transvalor - Presses des mines, ISBN: 978-2-35671-234-9.

[34] Archard, J. F. (1957). Elastic Deformation and the Laws of Friction. Proceedings of the Royal Society of London. Series A, Mathematical and Physical Sciences, 243 (1233), pp. 190-205.